\documentclass[twocolumn,showpacs,preprintnumbers,amsmath,amssymb,aps]{revtex4-1}
\usepackage{graphicx}
\usepackage{dcolumn}
\usepackage{bm}
\usepackage{float}
\usepackage{color}
\usepackage{soul}

\begin{document}

\title{Effects of Quantum Coherence on Work Statistics}

\author{Bao-Ming Xu$^{1}$}
\email{xubm2015@qfnu.edu.cn}
\author{Jian Zou$^{2}$}
\email{zoujian@bit.edu.cn}
\author{Li-Sha Guo$^{2}$}
\author{Xiang-Mu Kong$^{1}$}

\affiliation{$^{1}$School of Physics, Qufu Normal University, Qufu 273165, China}%
\affiliation{$^{2}$School of Physics, Beijing Institute of Technology, Beijing 100081, China}

\date{Submitted \today}

\begin{abstract}
In the conventional two-point measurement scheme of quantum thermodynamics, quantum coherence is destroyed by the first measurement. But as we know the coherence really plays an important role in the quantum thermodynamics process, and how to describe the work statistics for a quantum coherent process is still an open question. In this paper, we use the full counting statistics method to investigate the effects of quantum coherence on work statistics. First, we give a general discussion and show that for a quantum coherent process, work statistics is very different from that of the two-point measurement scheme, specifically the average work is increased or decreased and the work fluctuation can be decreased by quantum coherence, which strongly depends on the relative phase, the energy level structure and the external protocol. Then, we concretely consider a quenched 1-D transverse Ising model, and show that quantum coherence has a more significant influence on work statistics in the ferromagnetism regime compared with that in the paramagnetism regime, so that due to the presence of quantum coherence the work statistics can exhibit the critical phenonmenon even at high temperature.
\end{abstract}

\maketitle

\section{Introduction}

With recent experimental progress in the fabrication and manipulation of micro and nanoscale objects \cite{Greiner2002a,Greiner2002b,Kinoshita2006}, much attention has been given to understand the thermodynamics of small systems \cite{Lebowitz1961,Hill2013,Liphardt2005,Lucia2015,Chamberlin2015}. In such small systems the extensive thermodynamic quantities, such as work, heat and entropy production, might not be described by their average values alone, but their fluctuations should also be considered, just as done in the stochastic thermodynamics \cite{Seifert2008,Sekimoto2010,Jarzynski2011,Seifert2012,Broeck2015,Broeck2016,Jarzynski2017}. Stochastic thermodynamics has led to the discovery of various classical fluctuation theorems about work, heat and entropy production which connect microscopic dynamics with thermodynamic behaviors \cite{Jarzynski1997a,Crooks1999,Harris2007,Evans1993,Gallavotti1995,Jarzynski2000,Seifert2005,Ueda2010,Esposito2010,Kim2014,Gong2015,Seifert2016}.
For a small quantum system which obeys the laws of quantum mechanics, fluctuations are no longer just thermal in their origin but quantum as well. Now researchers are trying to extend the principles of thermodynamics to include quantum effects which should exist in small quantum systems \cite{Maruyama2009,Esposito2009,Campisi2011,Millen2016,Goold2016,Aberg2018}. It has been shown that quantum coherence \cite{Scully2003,Korzekwa2016,Li2014} and quantum correlations \cite{Llobet2015,Funo2013,Hovhannisyan2013} can be used to extract work. And based on the resource theory \cite{Horodecki2013b}, the second law of thermodynamics in the quantum regime was discussed from the perspective of quantum coherence (beyond free energy) \cite{Lostaglio2015b,Horodecki2015}. The nonequilibrium fluctuation relations in the closed quantum system hold unmodified \cite{Esposito2009,Campisi2011,Mukamel2003,Tasaki2000,Kurchan2000}. In the open quantum system, work, heat and entropy production were well defined based on the quantum trajectory approach (along the line of stochastic thermodynamics) \cite{Crooks2008b,Horowitz2012,Hekking2013,Leggio2013,Liu2014,Gong2016,Liu2016,Suomela2015,Breuer2003,Derezinski2008,Elouard2015,Deffner2011}, and their fluctuations were slightly modified \cite{Chernyak2004,Venkatesh2015,Cuetara2015,Campisi2009,Talkner2009,Crooks2008a}.

In general, to determine the work in the quantum regime one needs to perform two projective energy measurements at the beginning and the end of the external driving because the work is not an observable \cite{Talkner2007}. In quantum mechanics, the measurement will have a severe impact on the system dynamics and also on the statistics of work. It should be noted that the quantum effect is destroyed by the first measurement, and therefore the work fluctuation relation obtained by the two-point measurement scheme is not ``quantum" to some extent. How to describe the fluctuation of work for the quantum coherent process is still an open question. It has been proven that the measurement scheme that expecting the classical limit and obeying the first law of thermodynamics does not exist and this no-go result sheds light on the crucial roles of quantum measurement and quantum coherence \cite{Llobet2017}. Fortunately, a so-called full counting statistics (FCS) can describe the ¡°intrinsic¡± fluctuations of the system without any coupling to a measurement device \cite{Levitov1996,Nazarov2003,Clerk2011,Hofer2016}. Very recently, Solinas \textit{et al} used the FCS to investigate the full work distribution on a quantum system for arbitrary initial states \cite{Solinas2015}. Due to the presence of quantum coherence, the quasiprobability distributions of FCS can be negative \cite{Hofer2016}, and this pure quantum effects can be interpreted by the weak measurement theory \cite{Bednorz2010,Wei2008}. It has also been shown that the appearance of negativity of work quasidistribution is a direct signature of contextuality \cite{Lostaglio2017}.

In the present paper, we use the FCS method to investigate the effects of quantum coherence on work statistics including work quasidistribution, average work and work fluctuations. First, we give a general discussion. By dividing the initial state into coherent and incoherent parts, work is divided into the coherent work and the incoherent work, and their effects are carefully discriminated. The work statistics for a quantum coherent process is very different from that of a two-point measurement scheme; specifically, quantum coherence can increase or decrease average work and can decrease work fluctuation, which strongly depends on the relative phase, the energy level structure, and the external protocol. Then, we consider a quenched one-dimensional transverse quantum Ising model to concretely show these effects. The energy level structures of the Ising chain in different regimes (ferromagnetism and paramagnetism regimes) are very different, so that the responses of the Ising chain after the quench in different regimes are also different. After the quench, the response of the system in the ferromagnetism regime is much stronger than that in the paramagnetism regime. As a result, quantum coherence can significantly influence the work statistics in the ferromagnetism regime, but it does not in the paramagnetism regime. And we also find that in the presence of quantum coherence, the work statistics can exhibit the critical behavior at high temperature while it can-not without the coherence.

This paper is organized as follows: In the next section we briefly review some key concepts of work fluctuations based on the FCS method. In Sec. III we investigate the effects of quantum coherence on work statistics for arbitrary initial state and arbitrary external work protocol. In Sec. IV we investigate the influence of quantum coherence on the variation of free energy and the entropy production. A concrete quenched one-dimensional transverse quantum Ising model is considered in Sec. V. Finally, Sec. VI closes the paper with some concluding remarks.
\section{Work statistics based on the FCS method}

We begin by reviewing some key concepts of work statistics based on the FCS method in order to define the formalism that is used in the rest of this paper.

First, we demonstrate the general process considered in this paper: Consider a dynamical system described by a Hamiltonian $H_S(\lambda_t)$ that depends on an external work parameter $\lambda_t$, i.e., an externally controlled parameter. The Hamiltonian have a spectral decomposition $H_S(\lambda_t)=\sum_{n}\varepsilon_t^n|\psi^n_t\rangle\langle\psi^n_t|$. At the initial time $t=0$, the system-reservoir coupling is removed and a protocol is performed on the system with the work parameter being changed from its initial value $\lambda_{0}$ to the final value $\lambda_{\tau}$. After the protocol the system is in contact with a thermal equilibrium environment at temperature $T$. The environmental state can be described by Gibbs state $\rho^{G}_{B}=\exp(-\beta H_B)/Z_B$, with $H_B$ being the environment Hamiltonian, $\beta=1/T$ being the inverse of the temperature, and $Z_B=\mathrm{Tr}[\exp(-\beta H_B)]$ being the partition function of the environment. We assume that the system-environment coupling is weak. After a sufficiently long time, the system equilibrates with the thermal environment, and can be described by the Gibbs state $\rho^{G}_S(\lambda_\tau)=\exp(-\beta H_S(\lambda_\tau))/Z_S(\lambda_\tau)$ ($Z_S(\lambda_\tau)=\mathrm{Tr}[\exp(-\beta H_S(\lambda_\tau))]$ is the partition function of the system).

The work $W$ done by the external protocol is defined as the change of the internal energy between times $t=0$ and $t=\tau$ (before making contact with the environment). According to the FCS method, the characteristic function of the work done can be expressed as \cite{Solinas2015}
\begin{equation}\label{character}
\begin{split}
  \chi_{u}&=\mathrm{Tr}\bigl[e^{iuH_S(\lambda_{\tau})}U_S(\tau)e^{-i\frac{u}{2}H_S(\lambda_{0})}\rho_S(0)e^{-i\frac{u}{2}H_S(\lambda_{0})}U_S^{\dag}(\tau)\bigr],
  \\
  &=\sum_{lmn}e^{-iu\bigl(\varepsilon_\tau^l-(\varepsilon_0^m+\varepsilon_0^n)/2\bigr)}U_{lm}(\tau)\rho_{mn}(0)U^{\dag}_{nl}(\tau),
\end{split}
\end{equation}
where $U_S(\tau)\equiv\overleftarrow{T}\exp{\{-i\int_{0}^{\tau}H_S(\lambda_t)dt\}}$ is the time evolution operator, $\overleftarrow{T}$ is the time order operator, $U_{lm}(\tau)=\langle\psi_{\tau}^{l}|U_S(\tau)|\psi_{0}^{m}\rangle$, $\rho_S(0)$ is the initial state of the system, and $\rho_{mn}(0)=\langle\psi_{0}^{m}|\rho_S(0)|\psi_{0}^{n}\rangle$. All the moments of work can be determined by the standard way as $\langle W^{n}\rangle=(-i)^n\partial^{n}\chi_u/\partial u^{n}\arrowvert_{u=0}$. According to the characteristic function given by Eq. (\ref{character}), the moment of work distribution can be expressed as
\begin{equation}\label{work moment}
\begin{split}
&\langle W^{n}\rangle=\mathrm{Tr}\biggl[\bigl(U_S^{\dag}(\tau)H_S(\lambda_\tau)U_S(\tau)-H_S(\lambda_0)\bigr)^{n}\rho_S(0)\biggr] \\
  &=\sum_{lmm^{'}}\biggl(\varepsilon_\tau^l-\frac{\varepsilon_0^m+\varepsilon_0^{m^{'}}}{2}\biggr)^nU_{lm}(\tau)\rho_{mm^{'}}(0)U^{\dag}_{m^{'}l}(\tau).
\end{split}
\end{equation}
The work distribution (or quasidistribution) can be formally determined by the Fourier transform of the characteristic function $P(W)\equiv\int du e^{-iuW}\chi_{u}$. After the Fourier transform of the characteristic function given by Eq. (\ref{character}), the work distribution (quasidistribution) can be expressed as \cite{Solinas2016}
\begin{equation}\label{work_dis}
\begin{split}
  P(W)=\sum_{lmn}U_{lm}(\tau)\rho_{mn}(0)U^{\dag}_{nl}(\tau)
    \delta\biggl(W-\bigl(\varepsilon_{\tau}^{l}-\frac{\varepsilon_{0}^{m}+\varepsilon_{0}^{n}}{2}\bigr)\biggr).
  \end{split}
\end{equation}

If there is no quantum coherence in the initial state, i.e., all the off-diagonal elements of the density matrix (in the eigenbasis of initial Hamiltonian $H_S(\lambda_0)$) are zero, the work distribution can be simplified as $P(W)=\sum_{m,n}P^{m}_0P^{n|m}_\tau\delta(W-(\varepsilon_\tau^n-\varepsilon_0^m))$, with $P^{m}_0=\langle\psi^m_0|\rho_S(0)|\psi^m_0\rangle$ and $P^{n|m}_\tau=|\langle \psi^m_0|U_S(\tau)|\psi^n_\tau\rangle|^2$, which is just the result of the conventional two-point measurement scheme. Then, $P^{m}_0$ can be explained as the probability that the first measurement projects onto $|\psi^m_0\rangle$ at $t=0$, and $P^{n|m}_\tau$ is the conditional probability (conditioned on the first measurement result $|\psi^m_0\rangle$) that the second measurement obtains $\varepsilon_\tau^n$ at $t=\tau$ (before making contact with the environment). When the system is initially in the thermal state, i.e., $\rho_S(0)=\rho^{G}_S(\lambda_0)\equiv\exp(-\beta H_S(\lambda_0))/Z_S(\lambda_0)$, according to Eq. (\ref{character}) with $u=i\beta$ (because the characteristic function can be defined as $\chi_{u}\equiv\int dW e^{iuW}P(W)$), the well-known Jarzynski equality
\begin{equation}\label{Jarzynski equality}
  \langle e^{-\beta W}\rangle=\frac{Z_S(\lambda_\tau)}{Z_S(\lambda_0)}=e^{-\beta\Delta F}
\end{equation}
is obtained, where $\Delta F=T\ln Z_S(\lambda_\tau)/Z_S(\lambda_0)$ is the variation of the Helmholtz free energy. Remarkably, for the initial thermal state, the work fluctuation is solely determined by the equilibrium free energy difference $\Delta F$, but is independent of both the path wthere the work parameter is switched from $\lambda_0$ to $\lambda_\tau$ and the rate at which the parameter is switched along the path (i.e., independent of the nonequilibrium process determined by $U_S(\tau)$).

Not only can the FCS method recover the result of the two-point measurement scheme for an incoherent process, but it also has the unique advantage to investigate the effects of quantum coherence on the work statistics (quantum coherence is destroyed by the first measurement in the framework of the two-point measurement scheme). The work distribution (or quasidistribution) can be rewritten as $P(W)=\sum_{m,n}P^{m}_0P^{n|m}_\tau\delta(W-(\varepsilon_\tau^n-\varepsilon_0^m))+2\sum_{l,m>n}\mathrm{Re}[U_{lm}(\tau)\rho_{mn}(0)U^{\dag}_{nl}(\tau)]
\delta(W-(\varepsilon_{\tau}^{l}-(\varepsilon_{0}^{m}+\varepsilon_{0}^{n})/2))$.  Notably, $\sum_{l,m>n}\mathrm{Re}[U_{lm}(\tau)\rho_{mn}(0)U^{\dag}_{nl}(\tau)]=\mathrm{Tr}[U_S(\tau)(\rho_S(0)-\rho^{in}_S(0))U_S^{\dag}(0)]=0$, so that for some energy levels $\varepsilon^l_\tau$, $\varepsilon^m_0$, and $\varepsilon^n_0$, $\mathrm{Re}[U_{lm}(\tau)\rho_{mn}(0)U^{\dag}_{nl}(\tau)]$ can be negative, and the work quasidistribution can be negative, which strongly depend on the energy level structure ($\varepsilon^l_\tau$, $\varepsilon^m_0$, and $\varepsilon^n_0$) of the system and the external protocol ($U_S(\tau)$). The off-diagonal elements of the density matrix $\rho_{mn}(0)$ (i.e., quantum coherence) can be expressed as $\rho_{mn}(0)=|\rho_{mn}(0)|\exp{(i\phi_{mn})}$, with $\phi_{mn}$ being the relative phase of the initial state. Thus the negative quasidistribution also depends on the relative phase of the initial state. This negativity of $P(W)$ is a signature of the ¡°quantumness¡± of the work distribution and is destroyed by the first measurement in the two-point measurement scheme \cite{Solinas2016}. Using the FCS method to investigate the work statistics is just beginning and the effects of quantum coherence on the work statistics are not fully studied. Except for the negative quasidistribution, in the following we will comprehensively investigate the effects of quantum coherence on the work statistics, including average work and work fluctuation within the framework of FCS.

\section{Effect of quantum coherence on work statistics}
We divide the initial state into coherent and incoherent parts in the eigenbasis of initial Hamiltonian $H_S(\lambda_0)$, i.e.,
\begin{equation}\label{initial}
  \rho_S(0)=\rho^{in}_S(0)+\rho^{c}_S(0)
\end{equation}
with
\begin{equation}\label{}
  \rho^{in}_S(0)=\sum_mP^{m}_0|\psi^m_0\rangle\langle \psi^m_0|
\end{equation}
being the incoherence part of $\rho_S(0)$ and
\begin{equation}\label{}
  \rho^{c}_S(0)=\sum_{m\neq n}\rho_{mn}(0)|\psi^m_0\rangle\langle \psi^n_0|
\end{equation}
being the coherent part of the initial state in the eigenbasis of $H_S(\lambda_0)$.

\subsection{Average work}
Now we consider the average work $\langle W\rangle$. According to Eqs. (\ref{work moment}) and (\ref{initial}), the average work can be divided into two parts, i.e.,
\begin{equation}\label{AW}
  \langle W\rangle=\langle W\rangle^{in}+\langle W\rangle^{c}
\end{equation}
with
\begin{equation}\label{w_in}
  \langle W\rangle^{in}=\sum_{lm} P_0^m P^{l|m}_\tau(\varepsilon_\tau^l-\varepsilon_0^m)
\end{equation}
being the incoherent work and
\begin{equation}\label{W_c}
\begin{split}
\langle W\rangle^{c}=2\sum_{l,m>n}\varepsilon_\tau^l\mathrm{Re}[U_{lm}(\tau)\rho_{mn}(0)U^{\dag}_{nl}(\tau)]
\end{split}
\end{equation}
being the coherent work. In Eq. (\ref{W_c}), we have used $\sum_{lmm^{'}}(\varepsilon_0^m+\varepsilon_0^{m^{'}})U_{lm}(\tau)\rho_{mm^{'}}(0)U^{\dag}_{m^{'}l}(\tau)=\mathrm{Tr}[U_S(\tau)\{H_S(\lambda_0),\rho_S^c(0)\}U_S^\dag(\tau)]=0$. From Eqs. (\ref{w_in}) and (\ref{W_c}), we can see that the incoherent work is induced by the external protocol performed on the incoherent part of the initial state, i.e., $\langle W\rangle^{in}=\mathrm{Tr}[H_S(\lambda_\tau)\tilde{\rho}_S^{in}(\tau)-H_S(\lambda_0)\rho^{in}_S(0)]$, with $\tilde{\rho}^{in}(\tau)=U_S(\tau)\rho^{in}_S(0)U_S^{\dag}(\tau)$ being the evolution of the incoherent part of the initial state, and the coherent work is induced by the external protocol performed on the coherent part of the initial state, i.e., $\langle W\rangle^{c}=\mathrm{Tr}[H_S(\lambda_\tau)\tilde{\rho}^c(\tau)]$, with $\tilde{\rho}^{c}(\tau)=U_S(\tau)\rho^{c}_S(0)U_S^{\dag}(\tau)$ being the evolution of the coherent part of the initial state. Similarly, because $\mathrm{Re}[U_{lm}(\tau)\rho_{mn}(0)U^{\dag}_{nl}(\tau)]$ can be negative or positive depending on the external protocol, the relative phase of the initial state and the structure of energy level of system, the average work can be increased or decreased by quantum coherence, thus the quantum coherence can be used to improve the work extraction \cite{Scully2003,Korzekwa2016,Li2014}. Unlike the case of the two-point measurement scheme, the work done not only depends on the initial energy distribution $\rho^{in}_S(0)$ and final energy distribution, but also on the initial quantum coherence.

For further investigation, in the following, we rewrite the average work by using $\mathrm{Tr}[\rho_S(t)H_S(\lambda_{t})]=-T\mathrm{Tr}[\rho_S(t)\ln\rho^{G}_S(\lambda_t)]-T\ln Z_S(\lambda_t)$ and $S(\rho_1||\rho_2)=\mathrm{Tr}[\rho_1\ln\rho_1-\rho_1\ln\rho_2]$. The incoherent work can be expressed as
\begin{equation}\label{}
\begin{split}
    \langle W\rangle^{in}=&-T\ln Z_S(\lambda_\tau)/Z_S(\lambda_0)-TS(\rho^{in}_S(0)||\rho^G_S(\lambda_0)\\
    &+TS(\tilde{\rho}_S^{in}(\tau)||\rho^{G}_S(\lambda_\tau)).
\end{split}
\end{equation}
So that the incoherent work can be further divided into the external protocol independent part $\langle W\rangle^{in}_{indep}$ and the external protocol dependent part $\langle W\rangle^{in}_{dep}$, i.e.,
\begin{equation}\label{win}
  \langle W\rangle^{in}=\langle W\rangle^{in}_{indep}+\langle W\rangle^{in}_{dep}.
\end{equation}
The external protocol independent part is
\begin{equation}\label{expind}
\begin{split}
  \langle W\rangle^{in}_{indep}&=-T\ln\frac{Z_S(\lambda_\tau)}{Z_S(\lambda_0)}-TS\bigl(\rho^{in}_S(0)||\rho^G_S(\lambda_0)\bigr) \\
  &=-\sum_{m} P_{0}^m\varepsilon_0^m-T\sum_{m}P_{0}^m\ln P_{0}^m -T\ln Z_S(\lambda_\tau).
\end{split}
\end{equation}
From the first line of Eq. (\ref{expind}), it can be seen that the external protocol independent part is determined by the relative entropy between the incoherent part of the initial state and the equilibrium state with respect to the initial Hamiltonian $H_S(\lambda_0)$, and by the partition functions of the equilibrium states with respect to the initial and final Hamiltonians $H_S(\lambda_0)$ and $H_S(\lambda_\tau)$. The external protocol dependent part is
\begin{equation}\label{expd}
\begin{split}
  &\langle W\rangle^{in}_{dep}=TS\bigl(\tilde{\rho}_S^{in}(\tau)||\rho^{G}_S(\lambda_\tau)\bigr) \\
  &=\sum_{lm}P_0^m P^{l|m}_\tau\varepsilon_\tau^l+T\sum_{m}P_{0}^m\ln P_{0}^m+T\ln Z_S(\lambda_\tau).
\end{split}
\end{equation}
From the first line of Eq. (\ref{expd}), we can see that the external protocol dependent part is determined by the relative entropy between the evolution of the incoherent part of the initial state and the equilibrium state with respect to $H_S(\lambda_\tau)$.

The coherent work can also be expressed as
\begin{equation}\label{W_cc}
\begin{split}
\langle W\rangle^{c}&=TS\bigl(\rho_S(\tau)||\rho^{G}_S(\lambda_\tau)\bigr)
-TS\bigl(\tilde{\rho}_S^{in}(\tau)||\rho^{G}_S(\lambda_\tau)\bigr) \\
&-TS\bigl(\rho_S(\tau)||\tilde{\rho}_S^{in}(\tau)\bigr),
\end{split}
\end{equation}
where $\rho_S(\tau)=U_S(\tau)\rho_S(0)U_S^{\dag}(\tau)$ is the evolution of the initial state including the coherent part and the incoherent part. From Eq. (\ref{W_cc}), we can see that the coherent work is determined by the relative entropy between $\rho_S(\tau)$ and $\rho^{G}_S(\lambda_\tau)$ minus the sum of the relative entropy between $\tilde{\rho}_S^{in}(\tau)$ and $\rho^{G}_S(\lambda_\tau)$ and the relative entropy between $\rho_S(\tau)$ and $\tilde{\rho}_S^{in}(\tau)$.
\subsection{Work fluctuation}
According to Eqs. (\ref{work moment}) and (\ref{initial}), the second-order moment of work $\langle W^2\rangle$ can be divided into
\begin{equation}\label{}
  \langle W^2\rangle=\langle W^2\rangle^{in}+\langle W^2\rangle^{c}
\end{equation}
with
\begin{equation}\label{}
  \langle W^2\rangle^{in}=\sum_{lm}P_0^mP_\tau^{l|m}\bigl(\varepsilon_\tau^l-\varepsilon_0^m\bigr)^2
\end{equation}
being the incoherent part of $\langle W^2\rangle$ and
\begin{equation}\label{WC2}
\begin{split}
  &\langle W^2\rangle^{c} \\
  &=2\sum_{l,m>n}\bigl((\varepsilon_\tau^l)^2-\varepsilon_\tau^l(\varepsilon_0^m+\varepsilon_0^n)\bigr)\mathrm{Re}[U_{lm}(\tau)\rho_{mn}(0)U^{\dag}_{nl}(\tau)]
\end{split}
\end{equation}
being the coherent part of $\langle W^2\rangle$. In Eq. (\ref{WC2}), we have used $\sum_{lmm^{'}}(\varepsilon_0^m+\varepsilon_0^{m^{'}})^2U_{lm}(\tau)\rho_{mm^{'}}(0)U^{\dag}_{m^{'}l}(\tau)
=\mathrm{Tr}[U_S(\tau)H_S(\lambda_0)\rho_S^c(0)H_S(\lambda_0)U_S^\dag(\tau)]=0$. Because $\mathrm{Re}[U_{lm}(\tau)\rho_{mn}(0)U^{\dag}_{nl}(\tau)]$ can be positive or negative depending on the energy level structure of the system, the relative phase, and the external protocol, the second-order moment of work $\langle W^2\rangle$ can be increased or decreased by quantum coherence. Work fluctuations $\delta W^2\equiv\langle W^2\rangle-\langle W\rangle^2$ can be expressed as $\delta W^2=\delta W_{in}^2+\delta W_{c}^2-2\langle W\rangle^{in}\langle W\rangle^{c}$, where $\delta W_{in}^2=\langle W^2\rangle^{in}-(\langle W\rangle^{in})^2$ is the incoherent work fluctuation, $\delta W_c^2=\langle W^2\rangle^{c}-(\langle W\rangle^{c})^2$ is the coherent work fluctuation, and $2\langle W\rangle^{in}\langle W\rangle^{c}$ is the correlation between the incoherent work and coherent work. The incoherent work fluctuation $\delta W_{in}^2$ can be viewed as the work fluctuation obtained by the two-point measurement scheme on a coherent process where quantum coherence has been destroyed. The presence of quantum coherence can decrease the work fluctuation.

According to Eq. (\ref{character}) with $u=i\beta$, the work fluctuation relation for an arbitrary initial state can be expressed as
\begin{equation}\label{work statistics}
\begin{split}
&\langle e^{-\beta W}\rangle=\sum_{lm} P_0^m P^{l|m}_\tau e^{-\beta(\varepsilon_\tau^l-\varepsilon_0^m)} \\
&+2\sum_{l,m>n}e^{-\beta(\varepsilon_\tau^l-(\varepsilon_0^m+\varepsilon_0^n)/2)}\mathrm{Re}\bigl[U_{lm}(\tau)\rho_{mn}(0)U^{\dag}_{nl}(\tau)\bigr].
\end{split}
\end{equation}
From Eq. (\ref{work statistics}), it can be seen that the work fluctuation relation for an arbitrary initial state is no longer determined by the equilibrium states with respect to the initial and final Hamiltonians $H_S(\lambda_0)$ and $H_S(\lambda_\tau)$, i.e., it is no longer determined by $Z_S(\lambda_\tau)/Z_S(\lambda_0)$, but depends on the nonequilibrium process determined by $U_S(\tau)$. We divide the work fluctuation relation into two parts: the incoherent part which is induced by the incoherent part of the initial state (see the first line of Eq. (\ref{work statistics})) and the coherent part induced by the coherent part of the initial state (see the second line of Eq. (\ref{work statistics})). If the system is initially in equilibrium with the environment, i.e., $P_0^m=e^{-\beta\varepsilon_0^m}/Z_S(\lambda_0)$ and $\rho_{mn}=0$, the famous Jarzynski equality will be recovered. Because $\mathrm{Re}[U_{lm}(\tau)\rho_{mn}(0)U^{\dag}_{nl}(\tau)]$ can be negative, by manipulating the relative phase of the initial state and the external protocol, the presence of quantum coherence can decrease or increase $\langle e^{-\beta W}\rangle$ depending on the energy level structure of the system.

\section{Effect of quantum coherence on free energy and entropy production}

By dividing $\rho_S(t)$ (the density matrix of the system at any time $t$) into the coherent part and the incoherent part in the eigenbasis of Hamiltonian $H_S(\lambda_t)$, the free energy $F_t\equiv \mathrm{Tr}[H_S(\lambda_t)\rho_S(t)]+T\mathrm{Tr}[\rho_S(t)\ln\rho_S(t)]$ can be expressed as
\begin{equation}\label{free}
F_t=F^{in}_{t}+F^c_{t}
\end{equation}
with
\begin{equation}\label{freein}
F^{in}_t=-T\ln Z_S(\lambda_t)+TS(\rho^{in}_S(t)||\rho^{G}_S(\lambda_t))
\end{equation}
being the incoherent free energy, and
\begin{equation}\label{freec}
F^c_t=TS(\rho_S(t)||\rho^{in}_S(t))
\end{equation}
being the coherent free energy contributed by quantum coherence. $\rho^{in}_S(t)$ is the incoherent part of $\rho_S(t)$ just removing the coherence. It should be noted that the relative phase of $\rho_S(t)$ has no effect on the coherent free energy, which can be understood as follows. In general, the relative phase of $\rho_S(t)$ can be thought of as caused by the unitary evolution depending on Hamiltonian $H_S(\lambda_t)$, which does not influence the entropy of $\rho_S(t)$, and thus does not influence the relative entropy $S(\rho^{in}_S(t)||\rho^{G}_S(\lambda_t))$ and the coherent free energy.

In the weak system-environment coupling limit, the system after the thermalization can be described by the Gibbs state, $\rho^{G}_S(\lambda_\tau)=\exp(-\beta H_S(\lambda_\tau))/Z_S(\lambda_\tau)$ ($Z_S(\lambda_\tau)=\mathrm{Tr}[\exp(-\beta H_S(\lambda_\tau))]$), and the final free energy $F_\tau=-T\ln Z_S(\lambda_\tau)$.
Because the final equilibrium state is independent of the initial state, the coherence has no contribution to the final free energy $F_{\tau}$. According to Eq. (\ref{expind}) and  Eqs. (\ref{free})-(\ref{freec}), the variation of free energy $\Delta F\equiv F_\tau-F_0$ can be expressed as
\begin{equation}\label{DF}
\begin{split}
  \Delta F=\langle W\rangle^{in}_{indep}-F_0^c.
\end{split}
\end{equation}
From Eq. (\ref{DF}), it can be seen that the change of free energy consists of two parts: The first one is the incoherent work which is independent of the external protocol; the second one is the erasure of the initial coherent free energy $F_0^c$. Due to the erasure of quantum coherence, the entropy is increased by $S(\rho_S(0)||\rho^{in}_S(0))=\beta F_0^c\geq0$, so the amount of work extraction by an inverse process is decreased by $TS(\rho_S(0)||\rho^{in}_S(0))=F_0^c$. The free energy difference can also be written as $\Delta F=-T[\ln(Z_S(\lambda_\tau)/Z_S(\lambda_0))+S(\rho_S(0)||\rho^{G}_S(\lambda_0))]$ or, in other words, the free energy is rooted in the maximum work that can be extracted through the thermalization process.

Due to the irreversibility of the thermalization process, the work done on the system can not be fully stored as the free energy which is the maximum work being extracted by the inverse process, but a part of the work is used to produce entropy. The average entropy production is in the form of $\langle\Sigma\rangle=\beta\langle W_{\mathrm{irr}}\rangle$ with $\langle W_{\mathrm{irr}}\rangle\equiv\langle W\rangle-\Delta F$ being the average irreversible work. From Eqs. (\ref{AW}) and (\ref{DF}), the average entropy production can be expressed as
\begin{equation}\label{irr}
\begin{split}
\langle\Sigma\rangle=\beta\langle W\rangle^{in}_{dep}+\beta\langle W\rangle^{c}+\beta F^c_0.
\end{split}
\end{equation}
From Eq. (\ref{irr}), it can be seen that the coherent work and the incoherent work depending on the external process are used to produce entropy. And due to the erasure of the initial quantum coherence, the initial coherent free energy is dissipated, and the entropy (i.e., $S(\rho_S(0)||\rho^{in}_S(0))=\beta F^c_0$) is produced. The average entropy production can also be expressed as $\langle\Sigma\rangle=S(\rho_S(\tau)||\rho^{G}_S(\lambda_\tau))$, which means that the entropy production (or the irreversible work) is induced during the thermalization of the system state after the external protocol.

\section{Quenched 1-D Transverse Ising model}
A common way to drive an isolated quantum system out of equilibrium is by the so called sudden quench, where the Hamiltonian is abruptly changed from $H_S(\lambda_0)$ to $H_S(\lambda_\tau)$. Following the quantum quench, a number of fundamental questions on the nonequilibrium physics have aroused tremendous theoretical interest, ranging from the relationship between thermalization and integrability \cite{Polkovnikov2011} to the universality of defect generation at a quantum critical point \cite{Dziarmaga2005}. By treating the quench as a thermodynamic transformation, the characteristic function of work distribution was recognized to be the complex conjugate of the Loschmidt echo amplitude \cite{Silva2008}, such that the dynamical responses can be probed by the work done \cite{Wang2017}. Now we consider a quenched one-dimensional transverse quantum Ising model, to investigate the effects of quantum coherence on work statistics and concretely show the results in the above section. The quantum Ising model is regarded by Sachdev as one of two prototypical models to understand the quantum phase transition \cite{Sachdev2011}. The Hamiltonian of the quantum Ising model is
\begin{equation}\label{H0}
    H_S(\lambda)=-\sum_{j=1}^{N}\lambda\sigma_{j}^{x}+\sigma_{j}^{z}\sigma_{j+1}^{z},
\end{equation}
where $\lambda$ is a dimensionless parameter measuring the strength of the external field with respect to the spin-spin coupling. In this paper, we only consider $\lambda\geq0$ without loss of generality. $\sigma^{\alpha}_{j}$ $(\alpha=x,y,z)$ is the spin-1/2 Pauli operator acting on the $j$th spin and the periodic boundary conditions are imposed as $\sigma^{\alpha}_{N+1}=\sigma^{\alpha}_{1}$. Here we only consider that $N$ is even. For this model, the quantum phase transition takes place at the critical value $\lambda_c=1$ as the ordering of its ground state is discontinuous from a paramagnetic ($\lambda>1$) to a ferromagnetic ($\lambda<1$) phase. We call $\lambda>1$ the paramagnetism regime, and $\lambda<1$ the ferromagnetism regime, not only at low temperature but also at high temperature. After the diagonalization, the Hamiltonian (\ref{H0}) can be expressed as \cite{Sachdev2011}
\begin{equation}\label{H1}
    H_S(\lambda)=\sum_{k}\varepsilon_{k}(\gamma_{k}^{\dag}\gamma_{k}-\frac{1}{2}),
\end{equation}
where $\varepsilon_{k}=2\sqrt{\sin^{2}k+(\lambda-\cos k)^{2}}$, $\gamma_{k}$ is the Bogoliubov operator obeying the anticommutation $\{\gamma_{k},\gamma_{k'}^{\dag}\}=\delta_{kk'}$, $\{\gamma_{k},\gamma_{k'}\}=\{\gamma_{k}^{\dag},\gamma_{k'}^{\dag}\}=0$, $(\gamma_{k})^2=(\gamma_{k}^{\dag})^2=0$, and $k=\pm\frac{\pi}{N}(2n-1)$ with $n=1,\cdot\cdot\cdot,\frac{N}{2}$.
It should be noted that $\varepsilon_{k}=\varepsilon_{-k}>0$, and thus the Hamiltonian (\ref{H1}) can also be expressed as
\begin{equation}\label{}
    H_S(\lambda)=\sum_{k>0}\varepsilon_{k}(\gamma_{k}^{\dag}\gamma_{k}+\gamma_{-k}^{\dag}\gamma_{-k}-1).
\end{equation}

Here we consider a quench protocol where the external field is suddenly changed from $\lambda_{0}$ to $\lambda_{\tau}=\lambda_0+\delta_\lambda$, with $\delta_\lambda$ being the amplitude of the quench. After the quench protocol, the system is in contact with the environment at temperature $T$. The relation between pre- and post-quench Bogoliubov operators is \cite{Sachdev2011,Dorner2012}
\begin{equation}\label{RE}
\begin{split}
    &\gamma^{\tau}_{k}=\gamma^{0}_{k}\cos\frac{\Delta_{k}}{2}+\gamma^{0\dag}_{-k}\sin\frac{\Delta_{k}}{2}, \\
    &\gamma^{\tau}_{-k}=\gamma^{0}_{-k}\cos\frac{\Delta_{k}}{2}-\gamma^{0\dag}_{k}\sin\frac{\Delta_{k}}{2},
\end{split}
\end{equation}
where $\Delta_{k}=\theta^{\tau}_{k}-\theta^{0}_{k}$ and $\theta^{j}_{k}$ ($j=0,\tau$) is the Bogoliubov angle and can be defined by the relation
\begin{equation}\label{}
    e^{i\theta^{j}_{k}}=\frac{\lambda_{j}-e^{-ik}}{\sqrt{\sin^{2}k+\bigl(\lambda_{j}-\cos k\bigr)^{2}}}.
\end{equation}
It can be seen that $\theta^{j}_{-k}=-\theta^{j}_{k}$, and thus $\Delta_{-k}=-\Delta_{k}$, and from this the vacuum state in two representations is related by
\begin{equation}\label{basis}
  |0_{k}0_{-k}\rangle=(\cos\frac{\Delta_{k}}{2}+\sin\frac{\Delta_{k}}{2}\gamma^{\tau\dag}_{k}\gamma^{\tau\dag}_{-k})|\tilde{0}_{k}\tilde{0}_{-k}\rangle,
\end{equation}
where $|0_{k}0_{-k}\rangle$ is the vacuum state of $H_S(\lambda_0)$ and $|\tilde{0}_{k}\tilde{0}_{-k}\rangle$ is the vacuum state of $H_S(\lambda_\tau)$ for modes $k$ and $-k$.

In order to investigate the effects of quantum coherence, we consider that the spin chain is initially in a mixture of the Gibbs state and a so called coherent Gibbs state \cite{Lostaglio2015b,Kwon2018}:
\begin{equation}\label{initial state}
  \rho_S(0)=p|\Psi^G\rangle\langle\Psi^G|+(1-p)\rho^{G}_S(\lambda_{0}),~~~0\leqslant p\leqslant1,
\end{equation}
where
\begin{equation}\label{}
  |\Psi^G\rangle=\bigotimes_{k}\frac{1}{\sqrt{Z_{k}(\lambda_0)}}\biggl(e^{-\beta\varepsilon^{0}_{k}/4}|1_{k}\rangle
  +e^{\beta\varepsilon^{0}_{k}/4}e^{i\phi_k/2}|0_{k}\rangle\biggr)
\end{equation}
is the coherent Gibbs state with $Z_{k}(\lambda_0)=2\cosh(\beta\varepsilon^{0}_{k}/2)$ for mode $k$, and $\phi_k$ is the relative phase between the bases $|1_{k}\rangle$ and $|0_{k}\rangle$. It should be noted that $\prod_kZ_{k}(\lambda_0)=Z_S(\lambda_0)$ is the partition function of the Gibbs state $\rho^{G}_S(\lambda_{0})=1/Z_S(\lambda_0)\bigotimes_k(\exp(-\beta\varepsilon^{0}_{k}/2)|1_k\rangle\langle1_k|+\exp(\beta\varepsilon^{0}_{k}/2)|0_k\rangle\langle0_k|)$. If $p=1$, the spin chain is in the coherent Gibbs state $|\Psi^G\rangle$, but if $p=0$, the spin chain is in the thermal equilibrium state $\rho^{G}_S(\lambda_{0})$. The initial state can also be expressed as $\rho_S(0)=\rho^{G}_S(\lambda_{0})+p\bigotimes_{k>0}(e^{-i\phi_k}|1_k\rangle\langle0_{k}|+e^{i\phi_k}|0_k\rangle\langle1_k|)/Z_k(\lambda_0)$, where the incoherent part of the initial state is the thermal equilibrium state, i.e., $\rho^{in}_S(0)=\rho^{G}_S(\lambda_{0})$. It should be noted that $\rho_S(0)$ and $\rho^{G}_S(\lambda_{0})$ are energy indistinguishable because they have the same diagonal elements, and in this sense we call parameter $\beta=1/T$ in $\rho_S(0)$ the ``temperature" or ``effective temperature".

\subsection{Work statistics}
\subsubsection{work quasidistribution}
Because $\varepsilon_{k}=\varepsilon_{-k}$, we only consider modes $k>0$ and rewrite the initial state as $\rho_S(0)=1/Z_S(\lambda_0)\bigotimes_{k>0}(\exp(-\beta\varepsilon^{0}_{k})|1_k1_{-k}\rangle\langle1_k1_{-k}|+|1_k0_{-k}\rangle\langle1_k0_{-k}|+|0_k1_{-k}\rangle\langle0_k1_{-k}|
\small{+\exp(\beta\varepsilon^{0}_{k})|0_k0_{-k}\rangle\langle0_k0_{-k}|)}$ $+p\bigotimes_{k>0}(e^{-i\phi_k}|1_k1_{-k}\rangle\langle0_{k}0_{-k}|+|1_k0_{-k}\rangle\langle0_k1_{-k}|+\mathrm{h.c.})/Z^2_k(\lambda_0)$, where we assume that $\phi_{-k}=\phi_k$. According to Eq. (\ref{work_dis}), the work distribution for modes $k>0$ after the quench protocol is
\begin{equation}\label{k distri}
\begin{split}
&P(W_{k}=0)=\frac{2}{Z^{2}_{k}(\lambda_0)},\\
&P(W_{k}=\pm\varepsilon_{k}^{\tau}+\varepsilon_{k}^{0})
=\frac{e^{\beta\varepsilon^0_{k}}}{2Z^{2}_{k}(\lambda_0)}\bigl(1\mp\cos\Delta_k\bigr), \\
&P(W_{k}=\pm\varepsilon_{k}^{\tau}-\varepsilon_{k}^{0})
=\frac{e^{-\beta\varepsilon_{k}^0}}{2Z^{2}_{k}(\lambda_0)}\bigl(1\pm\cos\Delta_k\bigr), \\
&P(W_{k}=\pm\varepsilon^\tau_k)=\pm\frac{p\sin\Delta_k\cos\phi_k}{Z^{2}_{k}(\lambda_0)}.
\end{split}
\end{equation}
The first three terms, i.e., $P(W_{k}=0)$, $P(W_{k}=\pm\varepsilon_{k}^{\tau}+\varepsilon_{k}^{0})$, and $P(W_{k}=\pm\varepsilon_{k}^{\tau}-\varepsilon_{k}^{0})$, come from the incoherent part of the initial state, and can be obtained by the two-point measurement scheme. The last one, $P(W_{k}=\pm\varepsilon^\tau_k)$, comes from the coherent parts of the initial state, i.e., $p(e^{-i\phi_k}|1_k1_{-k}\rangle\langle0_{k}0_{-k}|+|1_k0_{-k}\rangle\langle0_k1_{-k}|+\mathrm{h.c.})/Z^2_k(\lambda_0)$. The influence of quantum coherence on the work done depends on the relative phase $\phi_k$ and the change of Bogoliubov angle $\Delta_k$. If $\phi_k=\pi/2$, the quantum coherence has no contribution to the work quasidistribution. Considering all the modes, the work quasidistribution can be expressed as
\begin{equation}\label{work distribution}
  P(W)=\sum_{\{\cdot\cdot\cdot W_{k}\cdot\cdot\cdot\}}\mathop{\biggl(\mathop{\prod_{k>0}P(W_{k})\biggr)}\delta\bigl(W-\small{\sum_{k>0}W_{k}}\bigr)},
\end{equation}
where $W_k=\{0,\varepsilon_k^\tau+\varepsilon_k^0,\varepsilon_k^0-\varepsilon_k^\tau,\varepsilon_k^\tau-\varepsilon_k^0,-\varepsilon_k^\tau-\varepsilon_k^0,\varepsilon^\tau_k,-\varepsilon^\tau_k       \}$.

Figures. 1(a1)-1(a3) show the work distribution at low temperature. In this case, the work distribution is independent of $p$ because the initial state is almost the ground state. In the paramagnetism regime $\lambda_0>1$, the work distribution is almost 1 for a definite amount of work, i.e., the amount of work performed by the external protocol is a definite value (see Fig. 1(a3)); now the work is completely described by the average work and the work fluctuation is of little importance. It should be noted that the width of the work distribution in the ferromagnetism regime (including the critical points) $\lambda_0\leq1$ is much wider (see Figs. 1(a1) and 1(a2)) than that in the paramagnetism regime $\lambda_0>1$; in other words, the work fluctuation in the ferromagnetism regime is more significant compared with that in the paramagnetism regime. This can be understood as follows: It is well known that the energy level structure of the Ising chain in different regimes (ferromagnetism and paramagnetism regimes) is very different, so that the responses of the system after the quench (the external field is suddenly changed from $\lambda_{0}$ to $\lambda_{\tau}$) in different regimes are also very different. In the paramagnetism regime, the spins are orientated randomly and are weakly affected by an externally applied magnetic field, so that the system response after suddenly changing the external field (i.e., the quench) is very weak. In the ferromagnetism regime, all the spins are oriented to the external field orientation, and after sudden change of the external field (i.e., the quench), the response of the system will be very strong. At low temperature, the initial state is nearly the ground state of the pre-quenched system. After the quench, the Hamiltonian is changed and the initial state is no longer the ground state of the post-quenched system. The quench protocol performs work on the system strongly depending on the response of the system, and the work distribution in the ferromagnetism regime (strong response) is wider than that in the paramagnetism regime (weak response). At the critical point $\lambda_0=1$, the gap between the ground state and the first excited state is vanished, the quench at the critical point reopens the gap and significantly changes the energy level structure, and the variation of Bogoliubov angle $\Delta_{k=0}$ is suddenly changed to $-\pi/4$, and thus the critical behavior can be observed. The width of work distribution for $\lambda_0\leq1$ is relatively wide, but that for $\lambda_0>1$ is relatively narrow, which means that the work distribution at low temperature can exhibit the phase transition or the phase transition from the paramagnetism regime to the ferromagnetism regime can widen the work distribution.

\begin{widetext}
\begin{center}
\includegraphics[width=16cm]{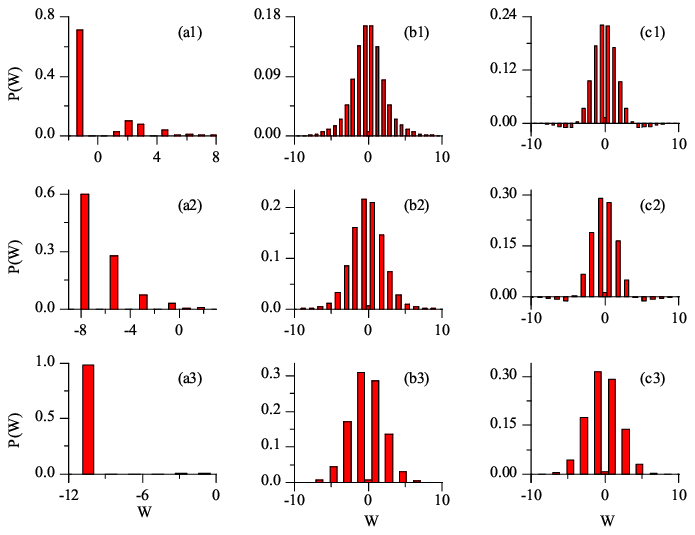}
\parbox{16cm}{\small{Fig. 1.} (Color online) The work distributions for various cases. (a1)-(a3) Low temperature $T=0.01$ and without coherence $p=0$; (b1)-(b3) high temperature $T=100$ and without coherence $p=0$; (c1)-(c3) high temperature $T=100$ and with coherence $p=1$. (a1),(b1),(c1) The ferromagnetism regime $\lambda_0=0$; (a2),(b2),(c2) the critical point $\lambda_0=1$; (a3),(b2),(c3) the paramagnetism regime $\lambda_0=2$. For all the panels, $\delta_\lambda=0.5$, $N=10$, and $\phi_k=\pi$.}
\end{center}
\end{widetext}

As the temperature increases, the work distribution will be widened and, in this case, the work fluctuation plays an important role. In the presence of quantum coherence, the work quasidistribution in the paramagnetism regime $\lambda_0>1$ is almost the same as that without considering quantum coherence (see Figs. 1(b3) and 1(c3)). However, the work quasidistribution in the ferromagnetism regime (including the critical point) $\lambda_0\leq1$ is very different from that without considering quantum coherence; more specifically, the work quasidistribution in the ferromagnetism regime (including the critical point) can be negative (see Figs. 1(c1) and 1(c2)). In other words, the appearance of negative distribution is the signature of the phase transition even at high temperature. Now we give a qualitative explanation for the appearance of negative distribution: The different effects of quantum coherence in different regimes (the ferromagnetism and paramagnetism regimes) mean that the influence of quantum coherence strongly depends on the energy level structure of the system (per the general discussion of Sec. III). We have shown that in the ferromagnetism regime (including the critical point) $\lambda_0\leq1$, the response of the system to the quench is strong, and then quantum coherence plays an important role. From the coherent work quasidistribution  $P(W_{k}=\pm\varepsilon^\tau_k)=\pm p\sin\Delta_k\cos\phi_k/Z^{2}_{k}(\lambda_0)$ (see Eq. (\ref{k distri})), it can be seen that the work quasidistribution can be negative.

\begin{center}
\includegraphics[width=8cm]{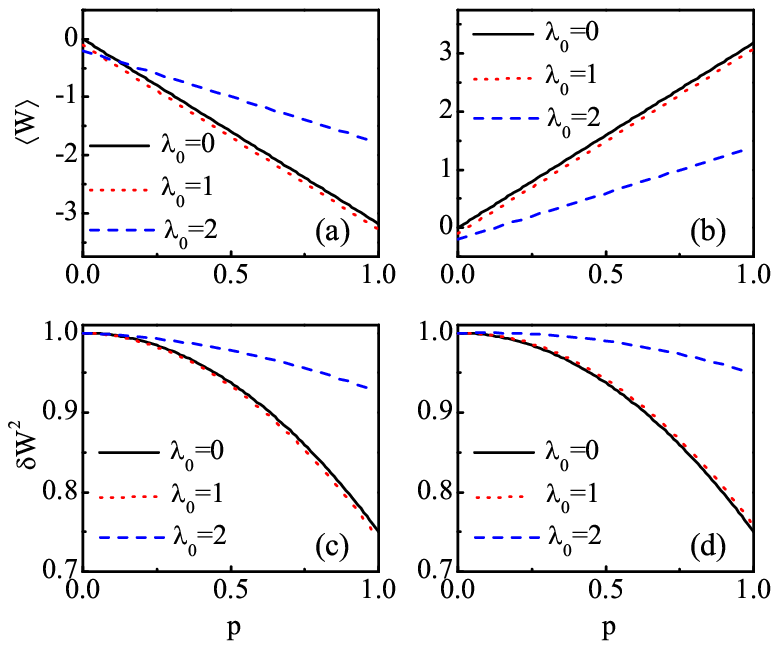}
\parbox{8cm}{\small{Fig. 2.} (Color online) The average work $\langle W\rangle$ and the work fluctuation $\delta W^2$ as functions of $p$ for different $\lambda_0$, and (a),(c) $\phi_k=0$ and (b),(d) $\pi$. For all the panels, $\delta_\lambda=0.1$, $T=100$, and $N=100$.}
\end{center}

\subsubsection{Average work and work fluctuation}

Now we investigate the average work (the first moment of work) and the work fluctuation (the second moment of work). Due to the incoherent part of the initial state $\rho^{in}_S(0)=\rho^{G}_S(\lambda_{0})$, the relative entropy $S(\rho^{in}_S(0)||\rho^{G}_S(\lambda_{0}))=0$, and thus the external protocol independent part of incoherent work $\langle W\rangle^{in}_{indep}=-T\ln(Z_S(\lambda_\tau)/Z_S(\lambda_0))$ (see Eq. (\ref{expind})). The external protocol dependent part of incoherent work $\langle W\rangle^{in}_{dep}=TS\bigl(\tilde{\rho}_S^{in}(\tau)||\rho^{G}_S(\lambda_\tau)\bigr)
=T\ln(Z_S(\lambda_\tau)/Z_S(\lambda_0))-\sum_{k>0}(\varepsilon^\tau_{k}\cos\Delta_{k}-\varepsilon^0_{k})\tanh(\beta\varepsilon^0_k/2)$ (see Eq. (\ref{expd})) and the coherent work $\langle W\rangle^{c}=2p\sum_{k>0}\varepsilon^\tau_k\sin\Delta_k\cos\phi_k/Z_{k}^2$ (see Eq. (\ref{W_c})). According to Eq. (\ref{AW}), the average work after the quench protocol can be expressed as
\begin{equation}\label{AW2}
    \langle W\rangle=\sum_{k>0}(\varepsilon^0_{k}-\varepsilon^\tau_{k}\cos\Delta_{k})\tanh\frac{\beta\varepsilon^0_k}{2}
    +\frac{2p\varepsilon^\tau_k\sin\Delta_k\cos\phi_k}{Z_{k}^2}.
\end{equation}

Figures. 2(a) and 2(b) show the effects of quantum coherence for $\phi_k=0$ and $\pi$ on the average work. It can be seen that the average work is decreased or increased with $p$ (i.e., quantum coherence) for the relative phase $\phi_k=0$ or $\pi$, respectively. And in the presence of quantum coherence ($p=1$), the average work is always negative or positive for $\phi_k=0$ or $\pi$, respectively. This can be understood as follows: At high temperature, the incoherent part of the initial state is almost the maximally mixed state, such that the average incoherent work is almost zero and the average coherent work plays a significant role; for the quench amplitude $\delta_{\lambda}>0$ that we considered, the change of Bogoliubov angle $\Delta_k<0$, and the quench protocol performs the negative or positive coherent work for $\phi_k=0$ or $\pi$.

After the quench protocol, the work fluctuation $\delta W^2=\langle W^2\rangle-\langle W\rangle^2$ can be expressed as
\begin{equation}\label{}
\begin{split}
  &\delta W^2=2\sum_{k>0}\frac{\bigl(\varepsilon_k^{\tau2}+\varepsilon_k^{02}-2\varepsilon_k^{\tau}\varepsilon_k^0\cos\Delta_k\bigr)\cosh\beta\varepsilon_k^0}{Z_k^2} \\
  &-\sum_{k>0}\biggl((\varepsilon^\tau_{k}\cos\Delta_{k}-\varepsilon^0_{k})\tanh\frac{\beta\varepsilon^0_k}{2}-\frac{2p\varepsilon^\tau_k\sin\Delta_k\cos\phi_k}{Z_{k}^2}\biggr)^2.
\end{split}
\end{equation}
Figures. 2(c) and 2(d) show the effects of quantum coherence on the work fluctuation. It can be seen that for an incoherent process ($p=0$), the work fluctuations for all $\lambda_0$ are the same, but for a coherent process ($p\neq0$), the work fluctuation decreases with $p$ (quantum coherence), i.e., the presence of quantum coherence can reduce the work fluctuation. In the presence of quantum coherence, the work fluctuations for $\phi_k=0$ and $\pi$ are almost the same, which is very different from the average work. It should be noted that the influences of quantum coherence on work and its fluctuation in the ferromagnetism regime (including the critical point) $\lambda_0\leq1$ are more significant than that in the paramagnetism regime $\lambda_0>1$. In other words, in the presence of quantum coherence, work and its fluctuation can exhibit the critical behavior at high temperature.

To show the critical behavior, we plot in Fig. 3 the average work and work fluctuation as the functions of $\lambda_0$ with and without considering quantum coherence. At low temperature, the work statistics is independent of $p$, so that we consider $p=0$ and plot the average work and work fluctuation as the functions of $\lambda_0$ in Figs. 3(a) and 3(c). It can be seen that the average work decreases with $\lambda_0$ and there is no singularity, while its derivative with respect to $\lambda_0$ has a singularity at the critical point $\lambda_0=1$ (see the inset in Fig. 3(a)). At low temperature, the work fluctuation in the ferromagnetism regime (including the critical points) $\lambda_0\leq1$ is relatively large and is independent of $\lambda_0$, but the work fluctuation in the paramagnetism regime $\lambda_0>1$ decreases with $\lambda_0$. In other words, at low temperature, not only the derivative of work fluctuation but also the work fluctuation itself show the critical behavior at the critical point $\lambda_0=1$. We have mentioned that after the quench, the response of the system in the ferromagnetism regime (including the critical point) $\lambda_0\leq1$ is much stronger than that in the paramagnetism regime, so that the work fluctuation in the ferromagnetism regime (including the critical point) is relatively large, and the critical behaviors of work and its fluctuations at the critical point $\lambda_0$ can be observed.

\begin{center}
\includegraphics[width=8cm]{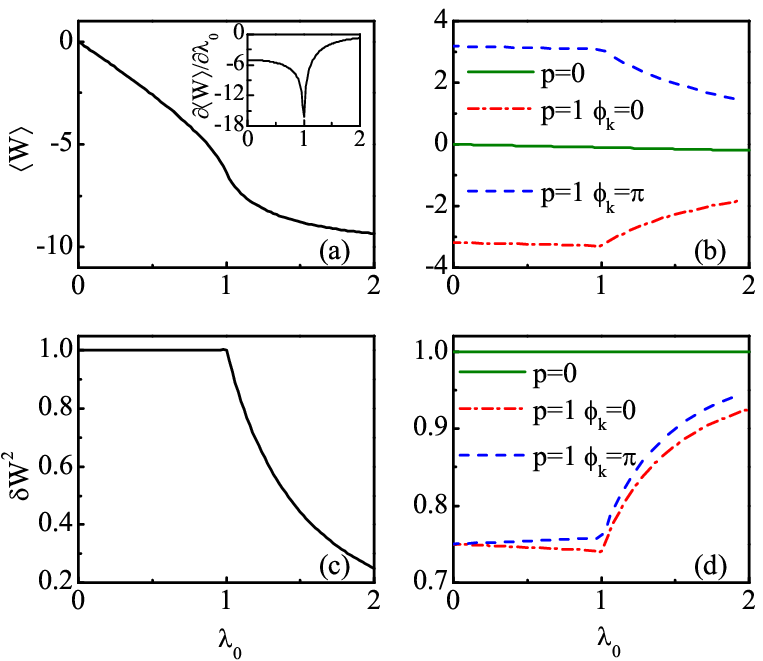}
\parbox{8cm}{\small{Fig. 3.} (Color online) Average work $\langle W\rangle$ and work fluctuation $\delta W^2$ as functions of $\lambda_0$ at (a),(c) $T=0.01$ and (b),(d) $T=100$. The inset in (a) is the derivative of average work with respect to $\lambda_0$ for $T=0.01$ and $p=0$. For all the panels, $\delta_\lambda=0.1$, $N=100$.}
\end{center}

At high temperature, the quantum coherence has significant effects on the work statistics. Considering $p=0$ and $p=1$, we plot the average work and work fluctuation as functions of $\lambda_0$ in Figs. 3(b) and 3(d). For the incoherent process ($p=0$), the work is almost independent of $\lambda_0$ because the initial state is almost the maximally mixed state which is independent of $\lambda_0$ and is not changed by the quench. For the coherent process (for example, $p=1$), the singularities of work (not only its derivative) and its fluctuation can be clearly observed at the critical point $\lambda_0=1$. The presence of quantum coherence, which makes the critical behavior of work and its fluctuation be observed at high temperature, can be understood as follows: The coherent Gibbs state is essentially a pure state which can be considered as the unitary transformation (rotation) of the ground state, i.e., $|\Psi^G\rangle=U(\beta)\bigotimes_k|0_k0_{-k}\rangle$, where $U(\beta)$ is the unitary transformation operator depending on the inverse of the temperature, and $U(\beta=\infty)=\mathbb{I}$, with $\mathbb{I}$ being the identity matrix. In other words, the initial state $|\Psi^G\rangle$ can be considered as the ground state of the unitary transformation of the original Hamiltonian (i.e., Ising model) $U(\beta)H_S(\lambda_0)U^\dag(\beta)$. This unitary transformation can not change the $Z_2$ symmetry of the original Hamiltonian, i.e., $U(\beta)H_S(\lambda_0)U^\dag(\beta)$ has the same symmetry as $H_S(\lambda_0)$, and thus the coherent Gibbs state $|\Psi^G\rangle$ can exhibit the critical behavior of the original Hamiltonian (i.e., Ising model). Quantum coherence will influence work and its fluctuations (and the work quasidistribution) depending on the energy level structure. We have shown that the responses of the system after the quench and thus the work performed by the quench in the ferromagnetism regime are different from that in the paramagnetism regime, so that quantum coherence has a more significant effect on the work and its fluctuation in the ferromagnetism regime than that in the paramagnetism regime. As a result, due to the presence of quantum coherence, the phase transition from ferromagnetism to paramagnetism can be observed by work and its fluctuation at high temperature. These critical behaviors are related to the appearance of the negative work quasidistribution.

\subsubsection{Work fluctuation relation}
From the Fourier transform of the work quasidistribution, the work fluctuation relation is obtained from Eq. (\ref{work statistics}),
\begin{equation}\label{C-F2}
\begin{split}
  \langle e^{-\beta W}\rangle=\ln\frac{Z_S(\lambda_\tau)}{Z_S(\lambda_0)}+\prod_{k>0}4p\sin\Delta_k\cos\phi_k\frac{\cosh\beta\varepsilon^\tau_k}{Z_k^2(\lambda_0)}.
\end{split}
\end{equation}
The second term of Eq. (\ref{C-F2}) comes from quantum coherence, which can increase or decrease $\langle e^{-\beta W}\rangle$ according to the relative phase $\phi_k$ and the Bogoliubov angle $\Delta_k$ (i.e., quench protocol) concretely. For $\Delta_k<0$, the relative phase $\phi_k\in[0,\pi/2)$ decreases $\langle e^{-\beta W}\rangle$ and, on the contrary, $\phi_k\in(\pi/2,\pi]$ increases $\langle e^{-\beta W}\rangle$. If there is no coherence in the initial state, i.e., $\rho_S(0)=\rho^{G}_S(\lambda_{0})$, the Jarzynski equality is recovered.

\subsection{Free energy and entropy production}

Now we will consider the free energy and the entropy production. The incoherent part of the initial state given by Eq. (\ref{initial state}) is the equilibrium state, i.e.,  $\rho^{in}_S(0)=\rho^{G}_S(\lambda_{0})$. According to Eq. (\ref{DF}), the coherent free energy of the initial state is
$F^c_0=T\sum_{k>0}2\Lambda_{k}^+\ln\Lambda_{k}^++2\Lambda_{k}^-\ln\Lambda_{k}^--\beta\varepsilon^0_{k}\tanh(\beta\varepsilon^0_k/2)+T\ln Z_S(\lambda_0)$, with $\Lambda^\pm_k=1/2\pm\sqrt{\sinh^2(\beta\varepsilon_k^0/2)+p^2}/(2\cosh\beta\varepsilon_k^0)$ being the eigenvalues of initial state $\rho_S(0)$, and the variation of free energy can be expressed as
\begin{equation}\label{DeltaF}
\begin{split}
    \Delta F&=T\sum_{k>0}\beta\varepsilon^0_{k}\tanh\frac{\beta\varepsilon^0_k}{2}-2\Lambda_{k}^+\ln\Lambda_{k}^+-2\Lambda_{k}^-\ln\Lambda_{k}^- \\
    &-T\ln Z_S(\lambda_\tau).
\end{split}
\end{equation}
From Eq. (\ref{DeltaF}), we can see that the variation of free energy is independent of the relative phase, which has been explained in the general discussion of Sec. III. The irreversible work after the quench protocol is
\begin{equation}\label{}
\begin{split}
\langle W_{irr}\rangle=&
    -\sum_{k>0}\varepsilon^\tau_{k}\cos\Delta_{k}\tanh\frac{\beta\varepsilon^0_k}{2}-\frac{2p\varepsilon^\tau_k\sin\Delta_k\cos\phi_k}{Z_{k}^2} \\
    &+\sum_{k>0}\Lambda_{k}^+\ln\Lambda_{k}^++2\Lambda_{k}^-\ln\Lambda_{k}^-+\ln Z_S(\lambda_\tau).
\end{split}
\end{equation}

\begin{center}
\includegraphics[width=8cm]{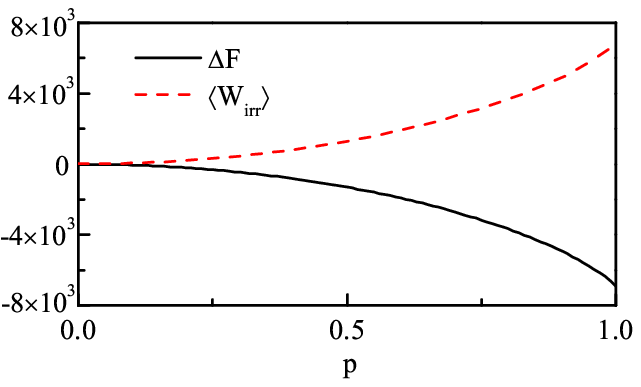}
\parbox{8cm}{\small{Fig. 4.} (Color online) The variation of free energy $\Delta F$ and the irreversible work $\langle W_{irr}\rangle$ as a function of $p$. $T=100$, $\lambda_0=0$, $\delta_\lambda=0.1$, $N=100$, $\phi_k=0$.}
\end{center}

The effects of quantum coherence on the variation of free energy and irreversible work (entropy production) for different $\lambda_0$ are similar and we consider $\lambda_0=0$ as an example and show the results in Fig. 4. It can be seen that the variation of free energy is dramatically reduced due to the erasure of the initial quantum coherence by the thermalization, and thus the irreversible work (entropy production) is dramatically increased. And the entropy is mainly produced by the erasure of the initial quantum coherence.

We also investigate the effects of $\lambda_0$ on the variation of free energy and the irreversible work (entropy production) for the incoherent ($p=0$) and coherent ($p\neq0$) processes. At low temperature, the free energy and irreversible work (entropy production) are independent of $p$, so we only consider $p=0$ and plot the results in Figs. 5(a) and 5(c). As expected, at low temperature, the singularities of free energy and irreversible work (entropy production) can be observed because the quench has a significant effect on the system in the ferromagnetism regime (including the critical point) while it has little effect on the system in the paramagnetism regime. The quantum criticality increases the irreversible entropy production (see Fig. 5(c)), which can be understood as follows \cite{Dorner2012}: The vanishing of the energy gap between the ground state and the first excited state makes it very difficult to drive the system across the critical region without exciting the system, and therefore a part of work is dissipated and entropy is produced. At high temperature, the variation of free energy and the irreversible work (entropy production) is almost independent of $\lambda_0$, whether or not quantum coherence is considered. The reason is that: for the incoherent process, i.e., $p=0$, the system at high temperature will almost be in the maximally mixed state which is independent of $\lambda_0$ and is not changed by the quench, so that the variation of free energy and the entropy production are independent of $\lambda_0$. For the quantum coherent process, i.e., $p\neq0$, the variation of free energy and the entropy production are mainly induced by the erasure of quantum coherence. And the quantum coherence at high temperature is almost independent of $\lambda_0$ (see Eq. (\ref{initial state})), so that the variations of free energy and the entropy production for the coherent process are also independent of $\lambda_0$.

\begin{center}
\includegraphics[width=8cm]{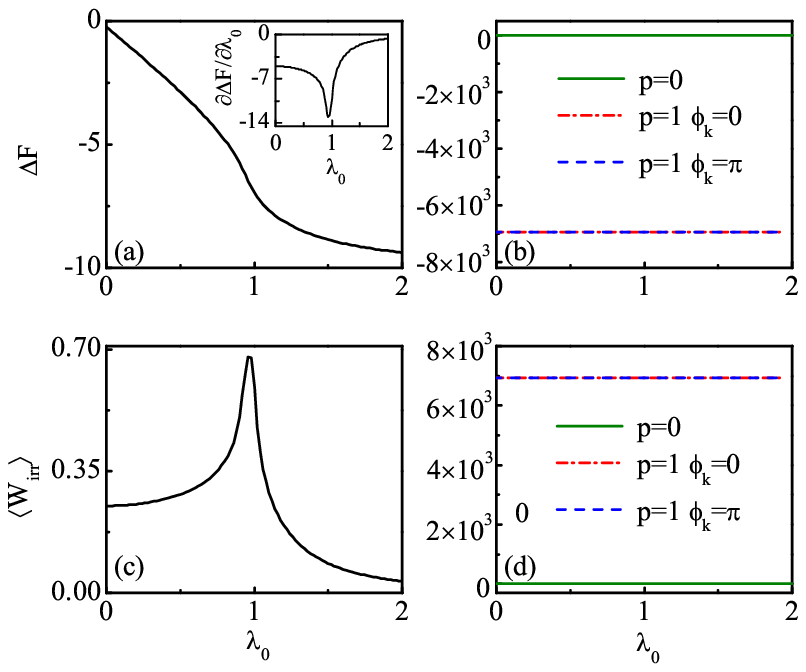}
\parbox{8cm}{\small{Fig. 5.} (Color online) The variation of free energy and irreversible work as functions of $\lambda_0$ for (a),(c) $T=0.01$ and (b),(d) $100$ . The inset in (a) is the derivative of free energy with respect to $\lambda_0$ for $T=0.01$ and $p=0$. It should be noted that the variations of free energy  in (b) for $p=1$, $\phi_k=0$ and $p=1$, $\phi_k=\pi$ coincide with each other, and similarly the irreversible work in (d) for $p=1$, $\phi_k=0$ and $p=1$, $\phi_k=\pi$ also coincide with each other. For all the panels, $\delta_\lambda=0.1$, $N=100$.}
\end{center}

\section{Conclusions}
In the spirit of the FCS, the effects of quantum coherence on the work statistics (including work quasidistribution, average work, and work fluctuation) are investigated in the present paper. First, we give a general discussion and show that for a quantum coherent process, work statistics is very different from that of the two-point measurement scheme, specifically, the average work is increased or decreased and the work fluctuation can be decreased, which strongly depends on the relative phase, the energy level structure, and the external protocol. Then, we concretely consider a quenched 1-D transverse Ising model, by which the analytical results can be obtained by means of a Jordan-Wigner transformation. We expect that the effects of quantum coherence on the work statistics in this simple model might be similar to those in more involved but less tractable models, so we can gain some insights into the quantum nonequilibrium fluctuations. Due to the presence of quantum coherence, work quasidistribution in the ferromagnetism regime can be negative; average work in the ferromagnetism regime is significantly decreased or increased by quantum coherence depending on relative phase, but work fluctuation in the ferromagnetism regime is only significantly decreased. In the paramagnetism regime, work statistics can be influenced by quantum coherence, but these influences are not significant compared with that in the ferromagnetism regime. These different effects of quantum coherence in different regimes (the ferromagnetism and paramagnetism regimes) mean that in the presence of quantum coherence, the work statistics can exhibit the critical behavior even at high temperature. The experimental measurement of work distribution requires the realization of an optical absorbtion experiment in a fully controllable setting. Recent proposals for the realization of quantum spin chains using bosonic atoms in optical lattices \cite{Duan2003} give a possible, concrete way to pursue this goal with the available experimental tools.

The quantum trajectory approach is also an important method to investigate the nonequilibrium fluctuation. Based on the quantum trajectory approach,  thermodynamic quantities for open quantum systems such as the work, heat, and entropy production were well defined, and their fluctuations has been widely investigated \cite{Crooks2008b,Horowitz2012,Hekking2013,Leggio2013,Liu2014,Gong2016,Liu2016,Suomela2015,Breuer2003,Derezinski2008,Elouard2015,Deffner2011}. In the quantum trajectory approach, the quantum system can initially stay at a superposition of two (or more) states, such that a coherent dynamics can be obtained. At a first glance, the quantum trajectory approach can be used to investigate fluctuations of work for quantum coherent process by considering an initial state with quantum coherence. However, in order to define the work along each quantum jump trajectory, one should know the energies at the beginning and the end of the dynamics, or the two-point energy measurement must be performed \cite{Liu2016}. In other words, if the system is initially in a superposition of two (or more) eigenstates, the work along each quantum jump trajectory can not be defined. If the thermodynamic quantity considered is a state function, just like the entropy, its fluctuation relation for the quantum coherent process can be investigated by the quantum trajectory approach. In the weak system-environment coupling limit, the entropy production can be defined as $\Sigma=\Delta S-\beta\Delta Q$, with $\Delta S$ being the change of the system entropy and $\Delta Q$ being the heat transferred from environment to the system. For the thermal equilibrium environment, the entropy production fluctuation $\langle\exp(-\Sigma)\rangle=1$ still holds for a quantum coherent process \cite{Deffner2011}.

\section{acknowledgement}
This work was supported by the National Natural Science Foundation of China (Grants No. 11705099, No. 11775019, No. 11675090, and No. 11504200).

\end{document}